\documentclass[12pt]{article}
\usepackage{amscd,amsmath,amssymb,amstext,amsthm,exscale,latexsym}
\newcommand {\al}   {\alpha}       \newcommand {\bt}  {\beta}
\newcommand {\g }   {\gamma}       \newcommand {\G }  {\Gamma}
\newcommand {\dl}   {\delta}       \newcommand {\e }  {\epsilon}
\newcommand {\z }   {\zeta}        
\newcommand {\ve}   {\varepsilon}  
\newcommand {\lm}   {\lambda}      
          
       \newcommand {\vr } {\varrho}
         
\newcommand {\vf }  {\varphi}      
         
\newcommand {\Lm}   {\Lambda}      \newcommand {\Om}  {\Omega}
       
\newcommand {\pl}   {\partial}     \newcommand {\nb}  {\nabla}
\renewcommand {\det}{{\sf\,det\,}}       \renewcommand {\dim}{{\sf\,dim\,}}
\renewcommand {\deg}{{\sf\,deg\,}}       \renewcommand {\div}{{\sf\,div\,}}
\newcommand   {\sign}{{\sf\,sign\,}}
\newcommand {\MM}  {{\mathbb M}}
\newcommand {\MS}  {{\mathbb S}}
\newcommand {\MU}  {{\mathbb U}}
\newcommand {\CM }  {{\cal M}}      \newcommand {\CN}  {{\cal N}}
\newcommand {\Se}  {{\textsc{e}}}   \newcommand {\Sh}  {{\textsc{h}}}
\newcommand {\Sm}  {{\textsc{m}}}   \newcommand {\Ss}  {{\textsc{s}}}
\begin{document}
\title     {Polynomial form of the Hilbert--Einstein action}
\author    {M. O. Katanaev
            \thanks{E-mail: katanaev@mi.ras.ru}\\ \\
            \sl Steklov Mathematical Institute,\\
            \sl Gubkin St. 8, Moscow, 119991, Russia}
\date      {21 March 2006}
\maketitle
\begin{abstract}
  Configuration space of general relativity is extended by inclusion
  of the determinant of the metric as a new independent variable.
  As the consequence the Hilbert--Einstein action takes a polynomial
  form.
\end{abstract}

Key words: Polynomial General Relativity

PACS: 04.20.-q
\vskip10mm

Let us consider a space-time $\MM$ of arbitrary dimension $\dim\MM=n\ge3$.
We do not deliberately restrict ourselves to the most interesting case
$n=4$ because gravity models in lower and higher number of dimensions
attracted much interest last years. For $n=2$ the Hilbert--Einstein
Lagrangian equals to a divergence and is not of interest here. The
construction in the present paper is valid for $n\ge3$ but the polynomiality
of the action is achieved for $n\ge4$. Local coordinates of the space-time
are denoted by $x^\al$, $\al=0,1,\dotsc,n-1$. Equations of motion for the
metric $g_{\al\bt}(x)$ in general relativity in the absence of matter
fields follow from the variational principle for the Hilbert--Einstein
action \cite{Hilber15R,Einste16R} with a cosmological constant $\Lm$
\begin{equation}                                        \label{ehelei}
  S_{\Sh\Se}=\int d^n x\sqrt g \left[R-(n-2)\Lm\right],~~~~g=|\det g_{\al\bt}|,
\end{equation}
where $R=R(g)$ is the scalar curvature for the metric $g_{\al\bt}$.
We use the following definitions of the curvature and Ricci tensors
and the scalar curvature
\begin{equation}                                       \label{ecurco}
\begin{split}
  R_{\al\bt\g}{}^\dl&=\pl_\al\G_{\bt\g}{}^\dl-\pl_\bt\G_{\al\g}{}^\dl
  -\G_{\al\g}{}^\e\G_{\bt\e}{}^\dl+\G_{\bt\g}{}^\e\G_{\al\e}{}^\dl,
\\
  R_{\al\bt}&=R_{\al\g\bt}{}^\g,~~~~R=g^{\al\bt}R_{\al\bt},
\end{split}
\end{equation}
where
\begin{equation}                                         \label{echrco}
  \G_{\al\bt}{}^\g=\frac12g^{\g\dl}(\pl_\al g_{\bt\dl}+\pl_\bt g_{\al\dl}
  -\pl_\dl g_{\al\bt})
\end{equation}
are Christoffel's symbols.

The coefficient in front of the cosmological constant is chosen in such a
way that the Einstein equations take the form
\begin{equation}                                        \label{eineql}
  R_{\al\bt}=\Lm g_{\al\bt}.
\end{equation}

The Hilbert--Einstein action (\ref{ehelei}) is not polynomial in metric
components $g_{\al\bt}$. First, it contains the nonpolynomial volume
element $\sqrt g$. Second, the expression for the scalar curvature contains
the inverse metric $g^{\al\bt}$ whose components are not polynomial in
$g_{\al\bt}$. Therefore the Hilbert--Einstein action in a perturbation theory
is represented by a very complicated infinite series, which is the main
difficulty in the analysis of the equations of motion and quantization.

Recently, a progress in approaching these problems has been made in the
framework of loop quantum gravity (see reviews \cite{Thiema03,AshLew04})
where the basic variables in the Hamiltonian formulation are not the
metric components and conjugate momenta but $\MS\MU(2)$-connection with
the corresponding momenta. In the present paper we follow a more
traditional way by extending the configuration space of metrics.

Coordinates of the configuration space $\CM$ in general relativity are
components of the metric $g_{\al\bt}(x)$. This space is infinite dimensional
and has
\begin{equation*}
  \frac{n(n+1)}2
\end{equation*}
coordinates at each point of space in a space-time $\MM$.

Let us consider different configuration space $\CN$ with coordinates
$\vr(x),k_{\al\bt}(x)$ assuming that $\vr>0$ and the matrix
$k_{\al\bt}$ is symmetric and nondegenerate in every point of a space-time:
$k_{\al\bt}=k_{\bt\al}$, $\det k_{\al\bt}\ne0$. We assume also that matrices
$g_{\al\bt}$ and $k_{\al\bt}$ have the same Lorentzian signature
\begin{equation*}
  \sign g_{\al\bt}=\sign k_{\al\bt}=(+\underbrace{-\dotsc-}_{n-1})
\end{equation*}
The dimensionality of the new configuration space at every point of space is
\begin{equation*}
  \frac{n(n+1)}2+1.
\end{equation*}

We define the subspace $\CM'$ in $\CN$ using the constraint
\begin{equation}                                        \label{emehed}
  \det k_{\al\bt}=\begin{cases}
    ~~1, & \text{if}~~\det g_{\al\bt}>0,~~~~\text{odd}~~n,
    \\-1, &\text{if}~~\det g_{\al\bt}<0,~~~\text{even}~~n. \end{cases}
\end{equation}
Then we can define the one-to-one correspondence between points of
the subspace $\CM'\subset\CN$ and the original configuration space $\CM$
\begin{equation}                                        \label{eghrel}
  g_{\al\bt}=\vr^2k_{\al\bt}.
\end{equation}
The inverse transformation is
\begin{equation}                                        \label{einmep}
  \vr=|\det g_{\al\bt}|^{1/{2n}},~~~~k_{\al\bt}=\vr^{-2}g_{\al\bt}.
\end{equation}
Hence we identify $\CM=\CM'$. The representation for the inverse metric is
$g^{\al\bt}=\vr^{-2}k^{\al\bt}$ where $k^{\al\bt}k_{\bt\g}=\dl^\al_\g$
follows from Eq.(\ref{eghrel}). The parameterization (\ref{eghrel}) for
$n=4$ was considered in \cite{Peres63}.

Let us make two important notes. First, components of the inverse
matrix $k^{\al\bt}$ are polynomial in $k_{\al\bt}$. In a general case
they are polynomials of order $n-1$ with respect to components $k_{\al\bt}$.
Second, the components $k_{\al\bt}$
are components of a second rank tensor density but not a tensor. Indeed,
we require Eq.(\ref{eghrel}) to be fulfilled in an arbitrary coordinate
system. Since the determinant of the metric $g$ is a scalar density of
weight $\deg g=-2$, then the matrix $k_{\al\bt}$ is the symmetrical tensor
density of second rank and weight $\deg k_{\al\bt}=2/n$, and the field $\vr$
is the scalar density of weight $\deg\vr=-1/n$. This means the following
transformations under coordinate changes $x^\al\rightarrow x^{\al'}(x)$
\begin{equation*}
  k_{\al'\bt'}=\pl_{\al'}x^\al\pl_{\bt'}x^\bt k_{\al\bt} J^{2/n},
  ~~~~\vr'=\vr J^{-1/n}.
\end{equation*}
where $J=\det\pl_\al x^{\al'}$ is the Jacobian of the coordinate transformation.
It means that the constraint (\ref{emehed}) is invariant under coordinate
transformations. Hence we call the matrix valued field $k_{\al\bt}(x)$
the {\em metric density}.

The explicit formula for the inverse metric density is
\begin{equation*}
  k^{\al\bt}=\frac1{(n-1)!}\hat\ve^{\al\g_2\dotsc\g_n}
  \hat\ve^{\bt\dl_2\dotsc\dl_n}k_{\g_2\dl_2}\dotsc k_{\g_n\dl_n},
\end{equation*}
where $\hat\ve^{\al_1\dotsc\al_n}$ is the totally antisymmetric tensor
density of weight $-1$ with unit components $|\hat\ve^{\al_1\dotsc\al_n}|=1$.
Here the summation is taken over $\g_2,\dotsc,\g_n$ and $\dl_2,\dotsc,\dl_n$,
and indices $\g_i$ and $\dl_i$, $i=2,\dotsc,n$ run over the whole range
$0,1,\dotsc,n-1$.
The given formula clearly shows that $k^{\al\bt}$ is polynomial in
$k_{\al\bt}$ of degree $n-1$ and vice versa.

Now we rewrite the Hilbert--Einstein action in terms of the new variables
$\vr,k_{\al\bt}$. Differentiations of Eq.(\ref{emehed}) yield the identities
needed for the calculations
\begin{align*}
  k^{\g\dl}\pl_\al k_{\g\dl}&=0,
\\
  k^{\g\dl}\pl^2_{\al\bt}k_{\g\dl}
  -k^{\g\dl}k^{\e\z}\pl_\al k_{\g\e}\pl_\bt k_{\dl\z}&=0.
\end{align*}
The expression for the scalar curvature is now easily found
\begin{equation}                                        \label{escurh}
  R=\vr^{-4}\left[\vr^2R^{(k)}+2(n-1)\vr\pl_\al(k^{\al\bt}\pl_\bt\vr)
  +(n-1)(n-4)\pl\vr^2\right],
\end{equation}
where we introduced the shorthand notation
\begin{equation*}
  \pl\vr^2=k^{\al\bt}\pl_\al\vr\pl_\bt\vr.
\end{equation*}
The ``scalar'' curvature $R^{(k)}$ for the metric density $k_{\al\bt}$
takes the surprisingly simple form
\begin{equation}                                        \label{escrha}
  R^{(k)}=\pl_{\al\bt}k^{\al\bt}
  +\frac12k^{\al\bt}\pl_\al k^{\g\dl}\pl_\g k_{\bt\dl}
  -\frac14k^{\al\bt}\pl_\al k^{\g\dl}\pl_\bt k_{\g\dl}.
\end{equation}
Note that this expression is polynomial in metric density $k_{\al\bt}$
as well as in its inverse $k^{\al\bt}$.

Let us make an important note. For a given metric density $k_{\al\bt}$ we can
formally construct Christoffel's symbols, curvature and Ricci ``tensors'' and
the ``scalar'' curvature using the original expressions (\ref{ecurco}) and
(\ref{echrco}). The corresponding Christoffel's symbols do not define a
connection on $\MM$, and curvature is not a tensor because $k_{\al\bt}$ is
now not a true tensor but a tensor density. For example, the scalar in
equation (\ref{escurh}) is not $R^{(k)}$ alone but the whole expression
in the right hand side. However, the group of general coordinate transformations
has a subgroup consisting of transformations with unit Jacobian. The
Christoffel's symbols $\G^{(k)}_{\al\bt}{}^\g$ for $k_{\al\bt}$ transform as
a connection coefficients with respect to this subgroup and the curvature
$R^{(k)}_{\al\bt\g}{}^\dl$ is a tensor.

Similar to general relativity, the second derivatives
$\pl^2_{\al\bt}k^{\al\bt}$ and $\pl_\al(k^{\al\bt}\pl_\bt\vr)$ can be
excluded from the action (\ref{ehelei}) by substraction of the boundary term
\begin{equation*}
  \pl_\al\left(\vr^{n-2}\pl_\bt k^{\al\bt}
  +2(n-1)\vr^{n-3}k^{\al\bt}\pl_\bt\vr\right).
\end{equation*}
This procedure does not alter the equations of motion and only simplifies
their derivation. The resulting action takes the form
\begin{equation}                                        \label{ehirid}
  S_{\Sh\Se}\overset{\div}{=}\int d^n xL,
\end{equation}
where
\begin{equation}                                        \label{ehieil}
\begin{split}
  L=\vr^{n-4}&\left[\frac12\vr^2k^{\al\bt}\pl_\al k^{\g\dl}\pl_\g k_{\bt\dl}
  -\frac14\vr^2 k^{\al\bt}\pl_\al k^{\g\dl}\pl_\bt k_{\g\dl}\right.
\\
  &~~~~\left.-(n-2)\vr\pl_\al k^{\al\bt}\pl_\bt\vr
  -(n-1)(n-2)\pl\vr^2-(n-2)\Lm\vr^4\phantom{\frac12}\right].
\end{split}
\end{equation}
For $n\ge4$ this Lagrangian is a polynomial in the fields $\vr$ and $k^{\al\bt}$
of order $n$ and $n+1$, respectively. The total degree of the polynomial
(\ref{ehieil}) is equal to $2n-1$.
The action (\ref{ehirid}) is equivalent to the original Hilbert--Einstein
action (\ref{ehelei}) and invariant under general coordinate transformations up to
boundary terms by construction. One has only to remember that the fields $\vr$ and
$k_{\al\bt}$ are not tensors but tensor densities.

The Lagrangian (\ref{ehieil}) is similar to the Lagrangian of the dilaton
gravity where the determinant of the metric plays the role of the dilaton.
It differs from the usual dilaton models by the presence of the cross
derivative term $\pl_\al k^{\al\bt}\pl_\bt\vr$. Besides, it contains less
number of independent fields because the constraint (\ref{emehed}) is imposed
on the metric density.

The extra constraint on the metric density (\ref{emehed}) can be
taken into account by addition of the constraint to the Lagrangian
\begin{equation*}
  L\rightarrow L+\lm(\det k_{\al\bt}\pm1),
\end{equation*}
where $\lm$ is a Lagrangian multiplier and $+$ or $-$ sign is taken for even
or odd dimensionality of the space-time, respectively. Remember that
$\det k_{\al\bt}$ is a true scalar, and therefore the constraint
$\det k_{\al\bt}\pm1=0$ has meaning in an arbitrary coordinate system.
This procedure is not necessary.
Variations of the metric density are restricted due to Eq.(\ref{emehed})
\begin{equation}                                         \label{ecomev}
  \dl\det k_{\al\bt}=\pm k^{\al\bt}\dl k_{\al\bt}
  =\mp\dl k^{\al\bt}k_{\al\bt}=0.
\end{equation}
The action (\ref{ehirid}) can be varied with respect to $k_{\al\bt}$ or
$k^{\al\bt}$ considering all components as independent and afterwards
taking the traceless part of the resulting equations.

The Euler--Lagrange equations for the action (\ref{ehirid}) are equivalent
to the Einstein equations (\ref{eineql}). The variation of the action
with respect to $\rho$ yields one equation
\begin{equation}                                           \label{etreie}
  2(n-1)\vr\square\vr+(n-1)(n-4)\pl\vr^2+R^{(k)}\vr^2=n\Lm\vr^4,
\end{equation}
which is proportional to the trace of the Einstein equations (\ref{eineql}).
Variation of (\ref{ehirid}) with respect to the metric density $k_{\al\bt}$
yields $n(n+1)/2-1$ equations
\begin{multline}                                           \label{etrles}
  \vr^2\left(R^{(k)}_{\al\bt}-\frac1nk_{\al\bt}R^{(k)}\right)+
\\
  +(n-2)(\vr\nb_\al\nb_\bt\vr-\nb_\al\vr\nb_\bt\vr)
  -\frac{n-2}nk_{\al\bt}(\vr\square\vr-\pl\vr^2)=0,
\end{multline}
where
\begin{equation}                                        \label{edefno}
\begin{split}
  \nb_\al\vr&=\pl_\al\vr,~~~~
  \square\vr=\pl_\al(k^{\al\bt}\pl_\bt\vr),
\\
  \nb_\al\nb_\bt\vr&=\pl^2_{\al\bt}\vr-\G^{(k)}_{\al\bt}{}^\g\pl_\g\vr,
\end{split}
\end{equation}
and Christoffel's symbols $\G^{(k)}_{\al\bt}{}^\g$ are computed for
the metric density $k_{\al\bt}$. If the action were varied without using the
constraint (\ref{emehed}) then one would get $n(n+1)/2$ equations similar to
Einstein's equations. However, the constraint restricts the variations by
equation (\ref{ecomev}) and therefore we must take only the traceless part
of the resulting equations. It is clear that equation (\ref{etrles}) has zero
trace and is proportional to the traceless part of Einstein's equations
(\ref{eineql}). Note that formal expressions for ``covariant'' derivatives for
tensors and tensor densities coincide as the consequence of Eq.(\ref{emehed})
because $\G^{(k)}_{\al\bt}{}^\bt=0$. The Ricci tensor in Eq.(\ref{etrles})
has the form
\begin{equation*}
\begin{split}
  R^{(k)}_{\al\bt}&=\frac12k^{\g\dl}(\pl^2_{\g\dl}k_{\al\bt}
  -\pl^2_{\al\g}k_{\bt\dl}-\pl^2_{\bt\g}k_{\al\dl})-\frac12\pl_\g k^{\g\dl}
  (\pl_\al k_{\bt\dl}+\pl_\bt k_{\al\dl}-\pl_\dl k_{\al\bt})
\\
  &-\frac14\pl_\al k_{\g\dl}\pl_\bt k^{\g\dl}-\frac12k^{\g\dl}k^{\e\z}
  (\pl_\g k_{\al\e}\pl_\dl k_{\bt\z}-\pl_\g k_{\al\e}\pl_\z k_{\bt\dl}).
\end{split}
\end{equation*}
Equations (\ref{etreie}) and (\ref{etrles}) are equivalent to the Einstein
equations in the following sense. For any solution $k_{\al\bt}$, $\rho$ of
equations (\ref{etreie}) and (\ref{etrles}) there is the unique metric
(\ref{eghrel}) and it satisfies the Einstein equations (\ref{eineql}).
Inverse, for any metric $g_{\al\bt}$ satisfying Einstein's equations
(\ref{eineql}) there are unique $k_{\al\bt}$, $\rho$ (\ref{einmep}), and
they satisfy equations (\ref{etreie}) and (\ref{etrles}).

Hence for $n\ge4$ the action (\ref{ehirid}), the Euler--Lagrange equations
(\ref{etreie}), (\ref{etrles}) and the constraint (\ref{emehed}) are polynomial
in the fields $\vr,k_{\al\bt}$. This important to our mind simplification is
achieved by extending the configuration space with the introduction
of additional field variable $\vr$. If the constraint (\ref{emehed}) is
solved with respect to one of the metric density components $k_{\al\bt}$
and the solution is substituted back into the action (\ref{ehirid}),
then polynomiality will be lost. Note that the introduction of new field
and constraint is not an extraordinary trick: the original metric
$g_{\al\bt}$ already contains nonphysical degrees of freedom to be excluded
from the theory by means of solution of the gauge conditions and constraints
present in general relativity. We increase simply the number of field
variables and constraints leaving the physical degrees of freedom untouched.

In the Hamiltonian framework the above procedure means the following.
The phase space corresponding to the variables $\vr,k_{\al\bt}$ is also
extended. There arises the additional constraint on the momenta
(the trace of the momenta corresponding to $k_{\al\bt}$ must be zero) which
together with the constraint (\ref{emehed}) form a pair of second class
constraints. However the total phase space will be no longer symplectic
but only Poisson manifold because the Poisson structure will
be degenerate. This question will be discussed elsewhere \cite{Katana06}.

Addition of a scalar $\vf(x)$ and electromagnetic $A_\al(x)$ fields
preserves polynomiality of the action and the Euler--Lagrange equations.
For the minimal coupling we have
\begin{align*}
  L_{\Ss}&=\sqrt g [g^{\al\bt}\pl_\al\vf\pl_\bt\vf-V(\vf)]
  =\vr^{n-2}[k^{\al\bt}\pl_\al\vf\pl_\bt\vf-\vr^2V(\vf)],
\\
  L_{\Se\Sm}&=\sqrt gg^{\al\bt}g^{\g\dl}F_{\al\g}F_{\bt\dl}
  =\vr^{n-4}k^{\al\bt}k^{\g\dl}F_{\al\g}F_{\bt\dl},
\end{align*}
where $V(\vf)$ is a potential for a scalar field including the mass term,
and $F_{\al\bt}=\pl_\al A_\bt-\pl_\bt A_\al$ is the electromagnetic field
strength.

In four-dimensional space-time Eq.(\ref{etreie}) is simplified
\begin{equation}                                        \label{efodtr}
  \square\vr+\frac16R^{(k)}\vr=\frac23\Lm\vr^3.
\end{equation}

Let us drop for a moment the constraint (\ref{emehed}) on the metric
density and consider $k_{\al\bt}$ and $\vr$ as a metric and a scalar field.
Then Eq.(\ref{efodtr}) is covariant with respect to conformal transformations
\begin{equation}                                        \label{econmt}
  \bar k_{\al\bt}=\Om^2k_{\al\bt},~~~~\bar\vr=\Om^{-1}\vr,
\end{equation}
where $\Om(x)>0$ is a two times differentiable function. It was considered
in \cite{Penros64} for $\Lm=0$ and in \cite{CaCoJa70} for $\Lm\ne0$.
The present approach is essentially different since the metric density
$k_{\al\bt}$ is subjected to the constraint (\ref{emehed}) which clearly
breaks conformal invariance. However the appearance of the factor $1/6$
is not accidental because the parameterization (\ref{eghrel}) coincides
with the conformal transformation (\ref{econmt}) in its form.

Gravity models for a metric with unit determinant were considered in physics
repeatedly. In generally covariant theories like in general relativity we
have an arbitrariness in choosing a coordinate system which can be utilized.
It is not difficult to prove that there is a coordinate system in a
neighborhood of an arbitrary point of space-time where the modulus of the
determinant of a metric equals unity $|\det g_{\al\bt}|=1$. Such coordinate
systems are defined up to coordinate transformations with unit Jacobian.
The condition $|\det g_{\al\bt}|=1$ simplifies essentially many formula,
in particular, the expression for the scalar curvature. This fact was used
by Einstein inventing general relativity \cite{Einste16R}. The "restricted"
gravity model based on the metric with unit determinant which is invariant
under general coordinate transformations preserving volume element attracted
interest some time ago [9--13].
\nocite{BijDam82,BucDra89A,HenTei89,Unruh89A,Kreuze90}

The choice of the fields $\vr,k^{\al\bt}$ as the basic variables is not the
unique way to achieve polynomiality of the Hilbert--Einstein action. If we
choose the inverse metric density
\begin{equation*}
  b^{\al\bt}=g^{\frac5{2(2n+1)}}g^{\al\bt}
\end{equation*}
as an independent variable, then the action will be a polynomial of degree
$2n+1$ without introduction of additional fields. For $n=4$ these variables
were considered in \cite{DeWitt67B}. However, the polynomiality will be lost
if we add the cosmological constant term in the action or minimally coupled
electrodynamics. The introduction of the additional field in (\ref{eghrel})
decreases the degree of polynomiality from $2n+1$ to $2n-1$ and survives
when interaction with matter is introduced.

Analogous but different parameterization  was used in \cite{LeoMla93} to
rewrite the Hilbert--Einstein action in a polynomial form.
The parameterization of the metric similar to (\ref{eghrel}) simplified
calculations in the renormalization analysis of $R^2$-gravity \cite{KalKaz97}.

More involved transformation of the phase space leading to polynomial
constraints in terms of complex variables was considered in \cite{Ashtek87}.

In the present case the polynomial action (\ref{ehirid}) is invariant
under general coordinate transformations up to boundary terms and is
equivalent to the Hilbert--Einstein action. It is very likely that the
proposed form of the action will be helpful in the construction of
quantum gravity.

The author is sincerely grateful to I.~V.~Volovich for the discussion
of the article.
This work is supported by RFBR (grant 05-01-00884) and the program of
support for leading academic schools (grant NSh-6705.2006.1).

\end{document}